# Pseudogap Behavior in Underdoped Cuprates

David Pines, Institute for Complex Adaptive Matter, University of California Office of the President, Physics Dept., UIUC, and Los Alamos National Laboratory

I review some of the experimental evidence and theoretical arguments that suggest that pseudogap matter is a new form of matter that coexists with coherent electron matter in the normal state and with superconducting matter below the superconducting transition temperature. I describe work in progress on a phenomenological two-fluid description of the evolution of pseudogap behavior that offers an explanation for the unexpectedly simple scaling behavior for the uniform magnetic susceptibility found in the underdoped cuprates and use this to propose a physical picture of the underdoped cuprates and to estimate the fraction of quasiparticles that become superconducting in underdoped superconductors.

I. Introduction

For those of us who predicted d-wave superconductivity in the cuprates, the discovery of d-wave symmetry has meant that the key question in the cuprates has shifted from "What is the mechanism for high Tc? (Our answer, " It is electronic and magnetic in origin") to fundamental questions about an unexpected new state of matter, pseudogap matter, that is found well above the superconducting transition temperature in underdoped cuprate superconductors. What is the physical origin of pseudogap behavior, and what is its onset temperature? What does experiment tell us about the resulting transformation of coherent electron states to pseudogap states and the description of pseudogap matter? What is the evidence that pseudogap behavior competes with superconductivity, and that this competition can lead to the existence of a quantum critical point?

In this talk, I review briefly some of the experimental evidence and theoretical arguments that suggest that pseudogap matter is a new form of matter that coexists with coherent electron matter in the normal state and with superconducting matter below the superconducting transition temperature. I describe work in progress on a phenomenological two-fluid description of the evolution of pseudogap behavior that offers an explanation for the unexpectedly simple scaling behavior found by Nakano, Johnston, et al for the uniform magnetic susceptibility in the underdoped cuprates and use their results to estimate the fraction of quasiparticles that become superconducting in underdoped superconductors.

II. The pseudogap state

Measurements of the uniform spin susceptibility of the 1-2-3 materials by Alloul and his colleagues in 1989 showed that in underdoped samples it possessed a maximum at a doping-dependent temperature, T*. The fall-off in that susceptibility below T* was attributed by Friedel to a gap in the quasiparticle spin spectrum, which he called a pseudogap. Subsequent experiments showed that a similar pseudogap phenomenon was present in all cuprate superconductors, while Johnston and Nakano showed in the 2-1-4 materials that when one plotted the ratio of the temperature dependent part of the uniform magnetic susceptibility to that at T*, as a function of T/T*, its temperature evolution displayed simple scaling behavior. Further insight into the pseudogap phenomenon came from NMR and ARPES experiments which suggested that in the underdoped cuprates an energy gap formed for the hot or antinodal quasiparticles (those near pi,0) in momentum space, while from their analysis of the antiferromagnetic component of the Cu spin-lattice relaxation rate and spin-echo decay rate, Barzykin and Pines argued that T* is the temperature at which the AF correlation length associated with the hot quasiparticles is of order 1 to 2 times the lattice spacing, a.

STM experiments suggested that in these materials at low temperatures one had intrinsic electronic spatial inhomogeneity. This conclusion is consistent with a proposal by Slichter on how one might reconcile the results of NMR experiments (a local probe) that seemed to require commensurate spin fluctuation peaks and INS experiments (a global probe) that show an incommensurate peak; Slichter proposed that one is observing discommensuration in the 2-1-4 (and possibly other cuprates), with regions of commensurate spin fluctuations being separated by small domains that are far less magnetic.

This past October at an Erice workshop, Seamus Davis presented convincing experimental evidence from his STM experiments on underdoped BSCCO that provide direct support for these scenarios (condmat0404005). He finds that pseudogap matter is made up of "hot" or anti-nodal quasiparticles and that it co-exists with superconducting matter at low temperatures, His results were supported by Shin-ichi Uchida in his Erice workshop presentation; along with Davis, Uchida argued that there are two distinct d-wave energy gaps, one associated with superconducting matter, one associated with pseudogap matter, and that the width of the distribution of the large doping dependent energy gaps in the underdoped materials reflects a range of densities in the regions of pseudogap matter. Uchida reported on experiments by Matsuda on the doping dependence of the range of measured energy gaps in the 2-1-4 materials that show this range narrowing as one increases the doping, with a very narrow distribution being measured for a doping level of 0.22. Matsuda's results suggest that for these materials, 0.22 marks the crossover from underdoped to overdoped materials. At Erice, Uchida also reported on experiments that indicate that the decrease in the condensation energy and the superfluid density seen as one goes toward increasingly underdoped materials may simply reflect the volume of the sample that is superconducting.

A simple physical picture that is consistent with the above results is that above T* one has coherent itinerant quasiparticle behavior over the entire Fermi surface, observed as an anomalous Fermi liquid. Below T* one loses that coherent behavior for a portion of the Fermi surface near the antinodes; the hot quasiparticles (those whose spin-fluctuation-induced interaction is strongest) found there enter the pseudogap state; its formation is characterized by a transfer of quasiparticle spectral weight from low to high frequencies that produces the decrease in the uniform spin susceptibility below T*. The remainder of the Fermi surface is largely unaffected.

In the pseudogap state at high temperatures one thus finds the coexistence of two distinct components; pseudogap matter and coherent quasiparticle matter; as the temperature is lowered it is the coherent quasiparticles that enter the superconducting state, while the pseudogapped quasiparticles are largely unaffected by the superconducting transition and do not participate in the superconducting behavior. One therefore expects two gaps in the superconducting state; a hot quasiparticle gap characterizing the pseudogap matter , with an energy ( of order 2 T*) that increases as the doping is reduced, and the d-wave gap of the superconducting matter (of order 3 kTc), that decreases as the doping is reduced.

What is the physical origin of the formation of pseudogap matter by the hot quasiparticles? It seems plausible that it is the increase in the strength of their antiferromagnetic correlations, as suggested by Barzykin and Pines. Viewed from the perspective of a hot quasiparticle, it is truly difficult for it to continue to be both itinerant and strongly antiferromagnetically correlated over distances somewhat greater than a lattice spacing. So a hot quasiparticle opts for the pseudogap state in which it is effectively localized and so finds no difficulty in becoming increasingly antiferromagnetically correlated as the temperature is lowered. Being effectively localized, it may also become spatially ordered. Quite importantly, Ali Yazdani 's group (Vershinin et al, Science 303, 1995, 2004) has found direct evidence from their STM experiments for such spatial ordering at temperatures above the superconducting transition in the BSCCO materials, while Seamus Davis also reported at Erice on evidence from his group's STM measurements on BSCCO at very low temperatures for the existence of periodic structures, presumably associated with regions of pseudogap matter that coexist there with superconductivity.

Pseudogap behavior is most easily identified at temperatures below T* but well above the superconducting transition temperature. A simple expression for the doping dependence of T* for the 2-1-4 materials that is consistent with the Nakano analysis and the Matsuda experiments is

$$T^* = 1250\,(1 - x/0.22)\text{ K}$$

where x is the hole doping level. To the extent that this extrapolation is correct for very low hole densities, it tells us that in the undoped material, the Mott insulator, at temperatures of order J, one enters the pseudogap state, out of which the Neel antiferromagnet forms at 400K or so. Since T* is of order J, it is plausible that it is a measure of the effective magnetic coupling between nearest neighbor Cu spins which might decrease as the hole density is increased. As one increases the doping above 0.02, one would then expect to find a coherent quasiparticle state coexisting with ordered pseudogap matter until one reaches doping levels above 0.06 or so, where at low temperatures the coherent quasiparticles become superconducting. The very recent optical experiments of the Basov group (Padilla et al, unpublished) provide striking evidence that this is the case. They see two components in the optical absorption, which are distinct at doping levels below 0.06: one is the Drude peak produced by coherent quasiparticles; the second is a mid-IR feature that it seems natural to identify with the pseudogap state. Rather surprisingly the coherent quasiparticle component of the normal state does not change its character appreciably with doping; the observed Drude peak corresponds to quasiparticles whose effective mass is independent of doping; only the height of the peak, proportional to the density of coherent quasiparticles, varies with doping throughout the entire underdoped region. Since the effective mass is doping independent, it is tempting to identify Basov's low frequency component as representing Fermi liquid islands of approximately constant density whose concentration increases with increasing doping.

A corollary of the above results is that one should expect isotope effects in the quasiparticle spectrum measured in the pseudogap state, since once localized, the hot quasiparticles can become strongly coupled to the lattice. (When not localized, the coupling of the hot quasiparticles to phonons is markedly reduced by vertex corrections associated with their magnetic coupling.) Any coupling of cold quasiparticles to the lattice would be very much smaller. These conclusions appear consistent with the ARPES results reported by Lanzara at this workshop. It leads me to predict that no isotope effect will be found for hot quasiparticles in overdoped materials.

As one increases doping, the above expression suggests that in the absence of superconductivity one would encounter a quantum critical point at a doping level of ~0.22. However this is likely not the case, since we have argued above that the low frequency behavior of the material associated with coherent quasiparticle matter is not affected by the emergence of the pseudogap state; the competition between the pseudogap state and the Fermi liquid state for the hot quasiparticles would therefore not give rise to quantum critical behavior. Put another way, the expected quantum critical point is hidden by the coherent quasiparticles whose behavior goes smoothly through the point in question.

It is worth remarking that the above phenomenology leads one to conclude that the marginal Fermi liquid behavior hypothesized for the optimally doped samples (x~0.16) does not reflect quantum critical behavior.

III. A Two-fluid Description of Emergent Pseudogap Behavior

The transfer of spectral weight from low frequencies to high frequencies that accompanies the formation of pseudogap matter is the inverse of process in heavy electron materials by which at some onset temperature the itinerant coherent heavy electron state emerges out of the local moments that make up the Kondo lattice. It has recently proved possible to develop a two-fluid description that describes this emergent coherent itinerant electron behavior and, in so doing, uncover fundamental scaling laws (Nakatsuji, Pines, and Fisk , PRL 92, 016401,2004); Curro, Young, Schmalian, and Pines (condmat 0402179). It is natural to inquire whether insight into the evolution of pseudogap matter can be obtained from the measured fall-off in $\chi(T)$ below $T^*$, by extending this two-fluid description to the underdoped cuprates, I thus assume that at temperatures below $T^*$ some fraction, $f(T)$, of the coherent hole states become part of pseudogap matter, and write the spin susceptibility as

$$\chi(T) = f(T) \chi_{PG}(T) + (1-f(T)) \chi_{COH}(T^*)$$

where $f(T)$ may be thought of as an order parameter characterizing the pseudogap state, $\chi_{PG}(T)$ describes the contribution of the pseudogapped quasiparticles to $\chi$, and I have made an assumption consistent with the Basov optical results-- that below $T^*$ the coherent quasiparticles cease to evolve so that the contribution of the coherent quasiparticles to $\chi$ is temperature independent, and given by their value at $T^*$. To the extent that the transfer of spectral weight to high frequencies for the hot quasiparticles is effective in eliminating their contribution to $\chi$, we may neglect $\chi_{PG}(T)$. We then find

$$\chi(T)/\chi(T^*) = 1 - f(T)$$

so that the fall-off in $\chi(T)$ below $T^*$ is explained as the loss of the coherent quasiparticle contribution to $\chi(T)$ and provides a direct measure of $f(T)$. If $T^*$ sets the scale for the temperature variation of $f$, i.e. $f(T/T^*)$, one obtains the scaling behavior of $\chi(T)$ proposed by Johnston and Nakano et al.

$1-f(Tc)$ then provides a direct measure of the fraction of the quasiparticles that are coherent and participate in the superconducting state. Assuming that the scaling behavior found for the 2-1-4 materials is universal, we may then use measured (or estimated) values of $T^*$ to obtain the following results

| Material | T*(K) | 1-f(Tc) |
| --- | --- | --- |
| 1-2-3 O 6.95 | 175 | 0.9 |
| 1-2-3 O 6.63 | 420 | 0.25 |
| 2-1-4 Sr 0.15 | 420 | 0.25 |

It will be interesting to see to what extent these results are consistent with the superconducting fraction obtained by other probes.

The much greater relative percentage of pseudogap matter in the 2-1-4 materials might then explain the differences in Tc between the 2-1-4 and 1-2-3 materials. If the superconducting fraction of the 2-1-4 materials behaves just like that in the quite overdoped region of the 1-2-3 materials where Tc increases from zero to, say, 40K, then since at this and lower 2-1-4 hole doping densities, pseudogap matter formation successfully competes with superconductivity, there would be no regions with a Tc greater than 40K.

IV. Outlook

The Davis experiments also suggest that in underdoped BSCCO materials, one has islands of local superconductivity embedded in a sea of pseudogap matter, a proposal that has been made by many people. The global result will then be granular d-wave superconductivity, brought about by Josephson coupling between d-wave islands whose hole density is that found near optimal doping, but whose relative fraction (volume) decreases as one decreases the overall doping level. This result is just what one would expect if it is the coherent Fermi liquid islands in the normal state that become superconducting. As noted earlier, a possible corollary of the Uchida results is that the increase in superfluid density with x, as one increases doping from the severely underdoped side, mainly reflects the increase in the relative number of superfluid regions whose average density is pretty much constant, and is of the order of that found near optimal doping. For this scenario to be consistent with a number of experiments that have been successfully interpreted as though one has bulk superconductivity, the Josephson coupling must be sufficiently strong that quasiparticles in the superconducting islands tunnel easily through pseudogap matter, so that at low frequencies there would be no appearance of inhomogeneity. D-wave superconductivity will help being this about, but an intimate relationship between pseudogap and superconducting matter may also be required.

The Hall effect measurements of Balakirev and Boebinger (BB) provide added support for a transition to pseudogap matter, since only that fraction of the Fermi surface that is superconducting would change its character at fields greater than that required to destroy superconductivity, and that fraction should change with doping. As one goes from the overdoped to the underdoped state at very low temperatures in the absence of superconductivity, one would expect to see a jump in the Hall effect whose magnitude would reflect the relative fraction of the quasiparticles that enter the pseudogap state. The existence of islands of superconductivity that, upon application of a strong enough magnetic field, become "cold" quasiparticle conducting islands embedded in insulating pseudogap matter, might provide a simple explanation of the log dependence on T found in the B B resistivity measurements in fields strong enough to destroy both granular sc and sc in the cold qp matter islands.

Clearly much more work needs to be done before the physical picture presented here is confirmed. Of particular interest is the issue of the order parameter that describes the pseudogap state which, as of this writing, is an open question.

V. Acknowledgements

I should like to thank Dmitri Basov, Seamus Davis, Juergen Haase, Alessandra Lanzara, Charlie Slichter, Shin-ichi Uchida, and Ali Yazdani for sharing their experimental results in advance of publication and for stimulating discussions concerning the interpretation of their results, and Artem Abanov, Andrey Chubukov and Steve Kivelson for stimulating discussions on these and related topics, This work has been supported in part by the Institute for Complex Adaptive Matter and the Department of Energy.